*systems*

# Visual Analysis of Nonlinear Dynamical Systems: Chaos, Fractals, Self-Similarity and the Limits of Prediction

**Geoff Boeing**

Department of City and Regional Planning, University of California, Berkeley, CA 94720, USA; gboeing@berkeley.edu; Tel.: +1-510-642-6000

**Abstract:** Nearly all nontrivial real-world systems are nonlinear dynamical systems. Chaos describes certain nonlinear dynamical systems that have a very sensitive dependence on initial conditions. Chaotic systems are always deterministic and may be very simple, yet they produce completely unpredictable and divergent behavior. Systems of nonlinear equations are difficult to solve analytically, and scientists have relied heavily on visual and qualitative approaches to discover and analyze the dynamics of nonlinearity. Indeed, few fields have drawn as heavily from visualization methods for their seminal innovations: from strange attractors, to bifurcation diagrams, to cobweb plots, to phase diagrams and embedding. Although the social sciences are increasingly studying these types of systems, seminal concepts remain murky or loosely adopted. This article has three aims. First, it argues for several visualization methods to critically analyze and understand the behavior of nonlinear dynamical systems. Second, it uses these visualizations to introduce the foundations of nonlinear dynamics, chaos, fractals, self-similarity and the limits of prediction. Finally, it presents Pynamical, an open-source Python package to easily visualize and explore nonlinear dynamical systems' behavior.

## 1. Introduction

Chaos theory is a branch of mathematics that deals with nonlinear dynamical systems. A system is simply a set of interacting components that form a larger whole. Nonlinear means that due to feedback or multiplicative effects between the components, the whole becomes something greater than the mere sum of its individual parts. Lastly, dynamical means the system changes over time based on its current state. Nearly every nontrivial real-world system is a nonlinear dynamical system. Chaotic systems are a type of nonlinear dynamical system that may contain very few interacting parts and may follow simple rules, but all have a very sensitive dependence on their initial conditions [1,2]. One might expect that any simple deterministic system would produce easily-predictable behavior. Yet, despite their deterministic simplicity, over time, these systems can produce wildly unpredictable, divergent and fractal (i.e., infinitely detailed and self-similar without ever actually repeating) behavior due to that sensitivity. Forecasting such systems' futures thus requires an impossible precision of measurement and computation. Chaos fundamentally indicates that there are limits to knowledge and prediction because some futures may be unknowable with any precision. Further, interventions into a system may have unpredictable outcomes even if the intervention is very minor, as tiny effects can compound (or be damped) nonlinearly over time.

Real-world chaotic and fractal systems span the spectrum from leaky faucets [3], to ferns [4], to heart rates [5–7], to cryptography [8,9]. Many scholars have studied the implications of nonlinearity, chaos and fractals for the social sciences, including sociology [10,11], urban studies [12–16], economics [17–21], architecture [22,23] and city planning [24–27]. One constant throughout the interdisciplinary history of nonlinear dynamical systems' study is that nonlinear systems are extremely difficult to solve analytically because they cannot be broken down into constituent parts, solved individually, then recombined as a solution. Scientists have instead relied heavily on visual and qualitative approaches, a perspective first developed by Henri Poincaré in the late 1800s, to discover and analyze the fascinating dynamics of nonlinearity [28,29]. Information visualization helps analysts detect and examine hidden structure in complex datasets [30]. In particular, few fields have drawn as heavily from visualization as nonlinear dynamics and chaos have for their pivotal discoveries, from Lorenz's first visualization of strange attractors [31], to May's groundbreaking bifurcation diagrams [32], to phase diagrams for discerning higher-dimensional hidden structures in data [33]. Such nonlinear analysis is particularly useful, yet underutilized for exploring time series [34,35].

These methods in turn have broad applicability to visual systems analysis. This article introduces nonlinearity and chaos interdisciplinarily through the methods of data visualization, using a logistic model to dissect the terminology, visualize pertinent features of chaos and fractals and discuss the wide-ranging implications for knowledge and prediction. It makes three primary contributions. First, it reviews and disseminates advanced visualization techniques for the qualitative analysis of nonlinear dynamical systems' behavior to an interdisciplinary body of systems scholars. Second, it provides a visual introduction to the salient concepts of nonlinearity and chaos to a scholarly audience. Although the social sciences are increasingly studying these types of systems, some of the seminal concepts remain murky or loosely adopted in the theoretical literature [36]. Most formal treatments of chaos and nonlinear dynamics in the scholarly literature are densely technical and geared toward an exclusive audience of mathematicians and physicists. For this article, rather, readers require only a familiarity with algebra. We thus do not cover the rigorous mathematical underpinnings of chaos and nonlinear dynamics, but the references throughout cite both the original foundational publications in this field, as well as recent scholarly developments. Interested readers will be well-rewarded in consulting these works. Third and finally, it presents Pynamical, a free open-source Python software package for the visual analysis of discrete nonlinear dynamical systems. Comparable tools usually must be developed from scratch or rely on expensive commercial software, such as MATLAB [37]. Pynamical provides a fast, simple, reusable and extensible new means for exploring system behavior.

The following section provides a background to the logistic map and the concepts of system dynamics and attractors. Then, we introduce several information visualization techniques to explore qualitative system behavior, bifurcations, the path to chaos, fractals and strange attractors. We investigate the difference between chaos and randomness before finally visualizing the famous butterfly effect and discussing its implications for scientific prediction. All of these models and visualizations are developed in Python using Pynamical; for readability, we reserve the technical details of its functionality for Appendix A.

## 2. Background and Model

Edward Lorenz, the father of chaos theory [38], once described chaos as "when the present determines the future, but the approximate present does not approximately determine the future" [39]. Lorenz first discovered chaos by accident while developing a simple mathematical model of atmospheric convection, using three ordinary differential equations [31]. He found that nearly indistinguishable initial conditions could produce completely divergent outcomes, rendering weather prediction impossible beyond a time horizon of about a fortnight [40].

How can this possibly happen with a simple deterministic system? We will explore an example using the logistic map, a model based on the common s-curve logistic function that shows how a population grows slowly, then rapidly, before tapering off as it reaches its environment's carrying capacity [41,42]. The logistic function uses a differential equation that treats time as continuous. The logistic map instead uses a difference equation to look at discrete time steps [43,44]. It is called the logistic map because it maps the population value at any time step to its value at the next time step: $x_{t+1} = r \cdot x_t \cdot (1 - x_t)$. This nonlinear equation defines the rules, or dynamics, of our system: $x$ represents the population at some time $t$, and $r$ represents the growth rate. Thus, the population level at any given time is a function of the growth rate parameter and the previous time step's population level. If the growth rate is set too low, the population will die out and go extinct. Higher growth rates might settle toward a stable value or fluctuate across a series of population booms and busts.

Chaos can manifest itself in both continuous (i.e., with dynamics defined by differential equations) and discrete (i.e., with dynamics defined by an iterated map) nonlinear dynamical systems. The logistic map is a simple, one-dimensional, discrete equation that produces chaos at certain growth rates. We will explore this in depth momentarily, but first, we use Pynamical to run the logistic model for 20 time steps (we will henceforth call these recursive iterations of the equation generations) for growth rate parameter values of 0.5, 1, 1.5, 2, 2.5, 3 and 3.5. Table 1 presents the results. The columns represent growth rates, and the rows represent generations. The model always starts with a population level of 0.5 and represents population as a ratio between zero (extinction) and one (the maximum carrying capacity of our system). If we trace down the column in Table 1 under a growth rate of 1.5, we see that the population level eventually settles toward a final value of 0.333 after several generations. In the column for a growth rate of two, we see an unchanging population level of 0.5 across every generation. This makes sense in the real world: if two parents produce two children, the overall population will neither grow nor shrink. Thus, a growth rate parameter value of two represents the replacement rate.

**Table 1.** Population values produced by the logistic map with 7 growth rate parameter values over 20 generations.

| Generation | $r = 0.5$ | $r = 1.0$ | $r = 1.5$ | $r = 2.0$ | $r = 2.5$ | $r = 3.0$ | $r = 3.5$ |
|---|---|---|---|---|---|---|---|
| 1 | 0.500 | 0.500 | 0.500 | 0.500 | 0.500 | 0.500 | 0.500 |
| 2 | 0.125 | 0.250 | 0.375 | 0.500 | 0.625 | 0.750 | 0.875 |
| 3 | 0.055 | 0.188 | 0.352 | 0.500 | 0.586 | 0.562 | 0.383 |
| 4 | 0.026 | 0.152 | 0.342 | 0.500 | 0.607 | 0.738 | 0.827 |
| 5 | 0.013 | 0.129 | 0.338 | 0.500 | 0.597 | 0.580 | 0.501 |
| 6 | 0.006 | 0.112 | 0.335 | 0.500 | 0.602 | 0.731 | 0.875 |
| 7 | 0.003 | 0.100 | 0.334 | 0.500 | 0.599 | 0.590 | 0.383 |
| 8 | 0.002 | 0.090 | 0.334 | 0.500 | 0.600 | 0.726 | 0.827 |
| 9 | 0.001 | 0.082 | 0.334 | 0.500 | 0.600 | 0.597 | 0.501 |
| 10 | 0.000 | 0.075 | 0.333 | 0.500 | 0.600 | 0.722 | 0.875 |
| 11 | 0.000 | 0.069 | 0.333 | 0.500 | 0.600 | 0.603 | 0.383 |
| 12 | 0.000 | 0.065 | 0.333 | 0.500 | 0.600 | 0.718 | 0.827 |
| 13 | 0.000 | 0.060 | 0.333 | 0.500 | 0.600 | 0.607 | 0.501 |
| 14 | 0.000 | 0.057 | 0.333 | 0.500 | 0.600 | 0.716 | 0.875 |
| 15 | 0.000 | 0.054 | 0.333 | 0.500 | 0.600 | 0.610 | 0.383 |
| 16 | 0.000 | 0.051 | 0.333 | 0.500 | 0.600 | 0.713 | 0.827 |
| 17 | 0.000 | 0.048 | 0.333 | 0.500 | 0.600 | 0.613 | 0.501 |
| 18 | 0.000 | 0.046 | 0.333 | 0.500 | 0.600 | 0.711 | 0.875 |
| 19 | 0.000 | 0.044 | 0.333 | 0.500 | 0.600 | 0.616 | 0.383 |
| 20 | 0.000 | 0.042 | 0.333 | 0.500 | 0.600 | 0.710 | 0.827 |

Figure 1 visualizes the resulting time series as a graph produced by Pynamical, with time on the *x*-axis and the system state on the *y*-axis. This graph visualizes how the population changes over time at different growth rates. For instance, the violet line for a growth rate of 0.5 quickly drops to zero: the population dies out. The teal line that represents a growth rate of two (the replacement rate) stays steady at a population level of 0.5. The growth rates of three and 3.5 are more interesting. While the green line for a growth rate of three seems to slowly converge toward a stable value, the yellow line for a growth rate of 3.5 just seems to repeatedly bounce around four different values.

An attractor is the value, or set of values, that a system settles toward over time. When the growth rate parameter is set to 0.5, the system has a fixed-point attractor at a population level of zero, as depicted by the violet line. In other words, the population value is drawn toward a stable equilibrium of zero over time as the model iterates: the logistic equation maps the value of a fixed-point attractor to itself. When the growth rate parameter is set to 3.5, the system oscillates between four values as depicted by the yellow line. This oscillating attractor is called a limit cycle. However, when we adjust the growth rate parameter beyond 3.57, we witness the onset of chaos. A chaotic system has a strange attractor, around which the system oscillates forever without ever repeating itself or settling into a steady state of behavior [45,46]. It never produces the same value twice, and its structure is fractal, meaning the same patterns exist at every scale no matter how much we zoom into it [47].

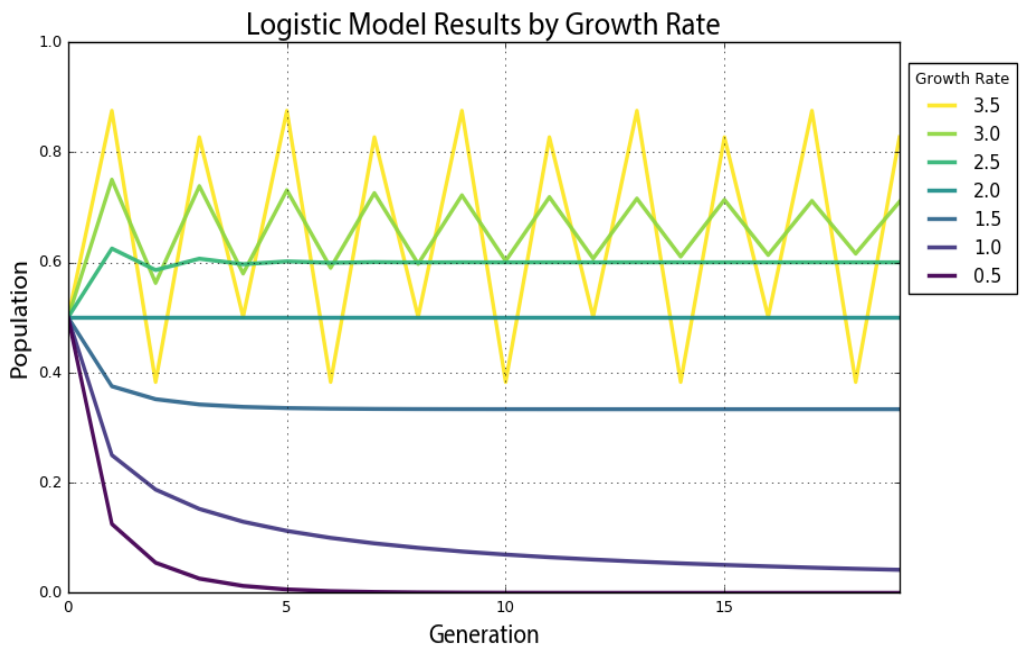

**Figure 1.** Time series graph of the logistic map with seven growth rate parameter values over 20 generations.

## 3. System Bifurcations

To show this more clearly, we run the logistic model again, this time for 200 generations across 1000 growth rate values between zero and four. When we produced the plot in Figure 1, we had only seven growth rates. This time, we have 1000, so we need to visualize the results in a different way to make them comprehensible, using a bifurcation diagram that visualizes a system's attractors as a function of some parameter [32,48,49]. The bifurcation diagram in Figure 2 represents 1000 discrete vertical slices, each corresponding to one of 1000 growth rate parameter values evenly spaced between zero and four. To produce each of these visual slices, Pynamical ran the model 200 times, then threw away the first 100 results, leaving just the final 100 generations for each growth rate. Each vertical slice thus visualizes the population values that the logistic map settles toward over time (i.e., the attractor) for that parameter value.

In Figure 2, we can see that for growth rates less than one, the system always eventually collapses to zero (extinction). For growth rates between one and three, the system always settles into an exact, stable population level. For instance, in the vertical slice above a growth rate of 2.5, there is only one population value represented (0.6), and it corresponds precisely to where the line for a growth rate of 2.5 settles in Figure 1's time graph. At this parameter value, the system's attractor is a fixed point at 0.6. However, for some growth rates, such as 3.9, the plot in Figure 2 shows 100 different values;

in other words, a different value for each of its 100 generations. Here, the system never settles into a fixed point or a limit cycle.

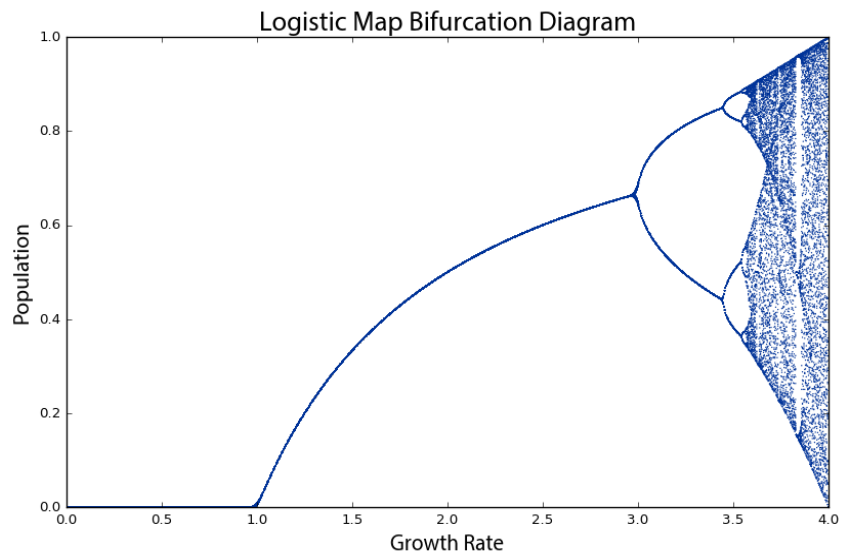

**Figure 2.** Bifurcation diagram of 100 generations of the logistic map for 1000 growth rate parameter values between zero and four. The vertical slice above each growth rate depicts the system's attractor at that rate.

Why is this visualization called a bifurcation diagram? If we zoom into the growth rates between 2.8 and 4 to see what is happening at a finer scale (Figure 3), the possible population values fork into two discrete paths at the vertical slice above a growth rate of three. At a growth rate of 3.2, the system oscillates exclusively between two population values: one around 0.5 and the other around 0.8. Thus, at that growth rate, applying the logistic map to one of these two population values yields the other. Just beyond a growth rate of 3.4, the diagram bifurcates again into four paths. This corresponds to the yellow line in Figure 1: when the growth rate parameter is set to 3.5, the system oscillates over four population values. These are periods, just like the period of a pendulum. At a growth rate of 3.2, the system has a period-two attractor. At a growth rate of 3.5, the system has a period-four attractor. Just beyond a growth rate of 3.5, it bifurcates again into eight population values. These consecutive bifurcations are phase transitions from one behavior, such as a fixed-point attractor, to a qualitatively different type of behavior, such as a period-two limit cycle attractor, as we vary the parameter value. Beyond a growth rate of 3.57, however, the bifurcations ramp up until the system is capable of eventually landing on any population value. This is known as the period-doubling path to chaos. As we adjust the growth rate parameter upwards, the logistic map will oscillate between two, then four, then eight, then 16, then 32 (and on and on to infinity) population values.

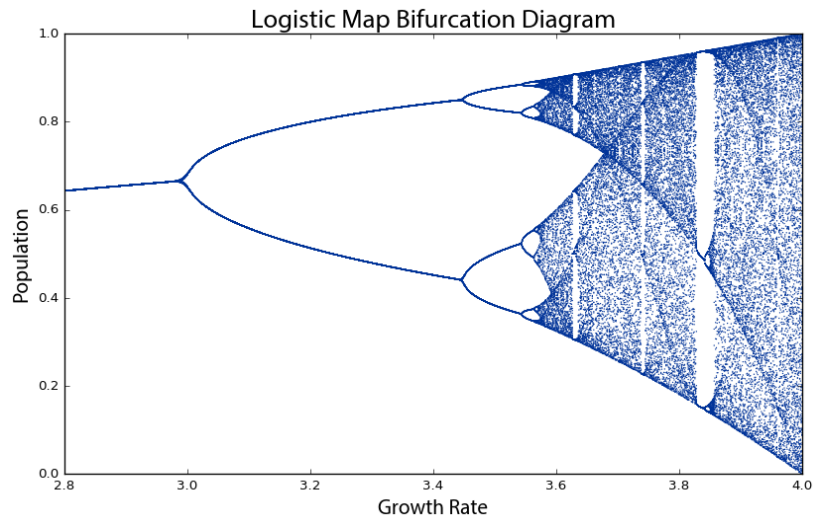

**Figure 3.** Bifurcation diagram of 100 generations of the logistic map for 1000 growth rate parameter values between 2.8 and 4. The vertical slice above each growth rate depicts the system's attractor at that rate.

By the time we reach a growth rate of 3.99, it has bifurcated so many times that the system now jumps, seemingly randomly, between all population values. We only say *seemingly* randomly because it is definitely not truly random. Rather, this model follows very simple deterministic rules yet produces apparent randomness due to its attractor having a period of infinite length. This is chaos: deterministic and aperiodic. If we zoom in again, to the narrow slice of growth rates between 3.7 and 3.9 (Figure 4), we begin to see the visceral beauty of chaos. Out of the noise emerge strange swirling patterns and thresholds on either side of which the system behaves very differently. For example, between the growth rates of 3.82 and 3.84, the system moves from chaos back into order, oscillating between just three population values: approximately 0.15, 0.55 and 0.95. However, then at growth rates beyond 3.86, it bifurcates again and returns to chaos. Indeed any one-dimensional system with a period-three cycle such as this at some parameter value is capable of chaotic behavior at other parameter values [50].

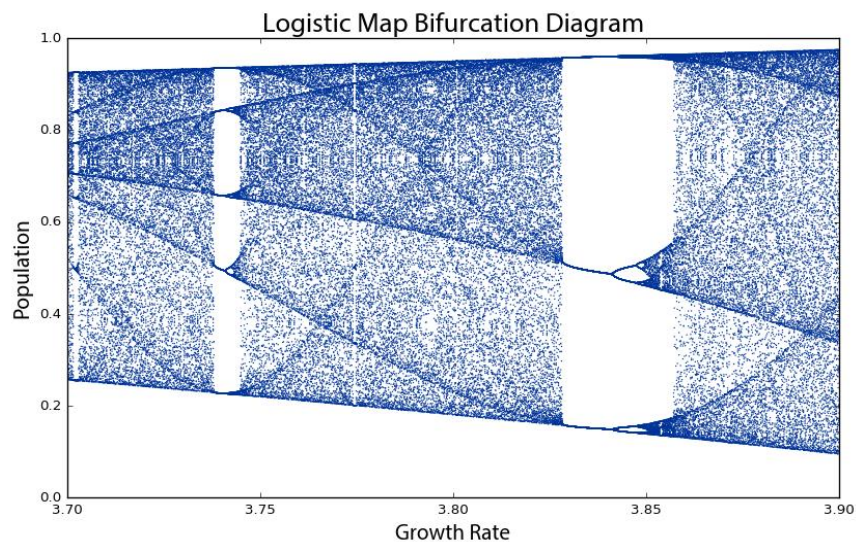

**Figure 4.** Bifurcation diagram of 100 generations of the logistic map for 1000 growth rate parameter values between 3.7 and 3.9. The system moves from order to chaos and back again as the growth rate is adjusted.

Universality refers to the phenomenon that very different systems can exhibit very similar behavior regardless of their underlying dynamics. It is commonly associated with Mitchell Feigenbaum's discovery that all systems that undergo this period-doubling path to chaos obey a mathematical constant [51,52]: the distance between consecutive bifurcations along the horizontal axis shrinks by a factor that asymptotically approaches 4.669, now known as Feigenbaum's constant [44]. Regardless of the system's specific dynamics, the ratio of the bifurcations on its road to chaos always obeys this constant.

## 4. Fractals and Strange Attractors

There is also a deep and universal connection between chaos and fractals [37]. In Figure 4, the bifurcations around a growth rate of 3.85 may look familiar. If we zoom in to the center one (Figure 5), we incredibly see the same structure that we saw earlier at the macro-level. In fact, if we keep zooming infinitely in to this visualization, we will continue seeing the same structure and patterns at finer and finer scales, forever. How can this possibly be? We mentioned earlier that chaotic systems have strange attractors and their structure can be characterized as fractal [53–55]. Fractals are shapes that are self-similar, meaning they have the same structure at every scale [56–58]. As we zoom in on them, we find smaller copies of the larger macro-structure. The bifurcation diagram (and thus, the attractor) of the logistic map is a fractal: at the fine scale in Figure 5, we see a tiny reiteration of the same bifurcations, chaos and limit cycles we saw in Figure 1's visualization of the full range of growth rates.

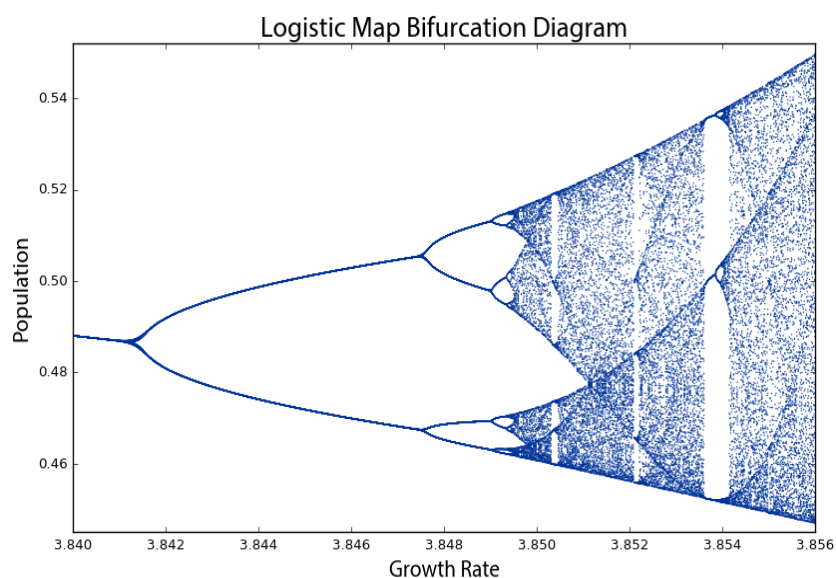

**Figure 5.** Bifurcation diagram of 100 generations of the logistic map for 1000 growth rate parameter values between 3.84 and 3.856. This is the same structure that we saw earlier at the macro-level in Figure 3, because chaotic systems' strange attractors are fractal.

Another way to visualize this nonlinear time series is with a phase diagram, using a method called state-space reconstruction through delay-coordinate embedding [35]. Simply put, this plots the system's value at generation $t + 1$ on the $y$-axis versus its value at $t$ on the $x$-axis [59], giving us another visual window into the qualitative behavior of the system. The clever insight of this phase diagram is that it embeds one-dimensional time series data from our logistic map into two-dimensional state space: an imaginary space that uses system variables as its dimensions [33,60,61]. Each point in state space is a system state, or in other words, a set of variable values. While traditional systems' analysis tends to focus on visualizing time series as in Figure 1, nonlinear dynamics tends to focus on visualizing these state spaces. Few real-world systems are fully observable, yet the dynamics in a properly reconstructed state space are identical to the true dynamics of the entire system [34].

In our case, the two variables are: (1) the population value at generation *t*; and (2) the value at *t* + 1. For example, with a growth rate of 3.5, the population value at Generation 1 is 0.5; the value at Generation 2 is 0.875; the value at Generation 3 is 0.383; and so forth (see Table 1). Therefore, our two-dimensional phase diagram will have (*x*, *y*) points at (0.5, 0.875) and (0.875, 0.383), and so on (Figure 6B). Remember that our model follows a simple deterministic rule, so if we know a certain generation's population value, we can easily determine the next generation's value. Like earlier, to produce these phase diagrams, Pynamical runs the logistic model for 200 generations and then discards the first 100 rows, to visualize only those values that the system settles toward over time.

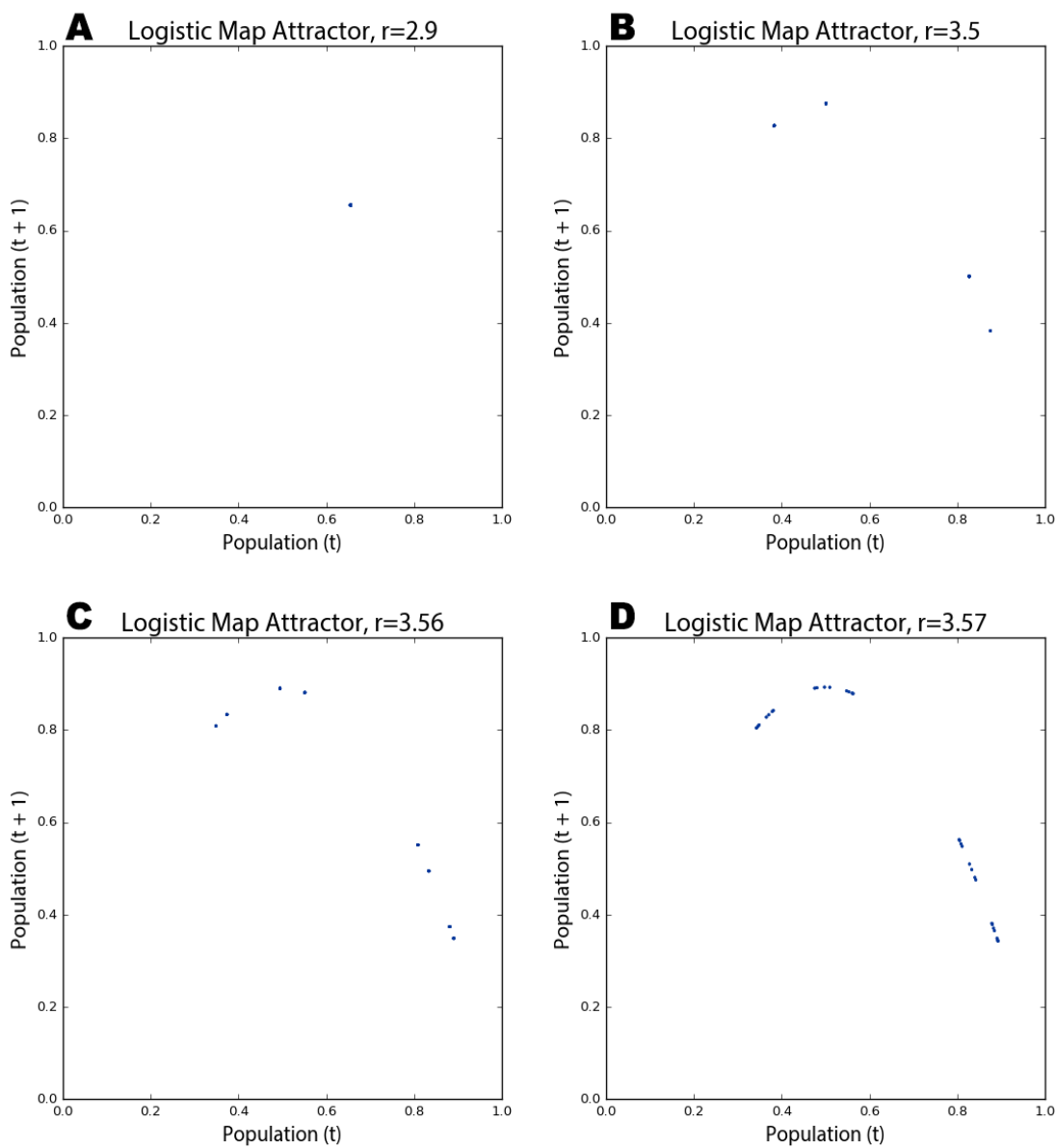

**Figure 6.** Phase diagrams of the logistic map over 200 generations for growth rate parameter values of: 2.9 (**A**); 3.5 (**B**); 3.56 (**C**); and 3.57 (**D**). When the parameter is set to 2.9, the model converges at a single fixed-point. When the parameter is set to 3.5 or higher, the model oscillates over four points, then eight, and on and on as it bifurcates.

In Figure 6A, the phase diagram shows that the logistic map homes in on a fixed-point attractor at 0.655 (on both axes) when the growth rate parameter is set to 2.9. This corresponds to the vertical slice above the *x*-axis value of 2.9 in the bifurcation diagram in Figure 2. Figure 6B depicts a period-four limit cycle attractor: when the growth rate is set to 3.5, the logistic map oscillates over four points, as shown in this phase diagram (and in Figures 1 and 2). If we adjust the growth rate parameter up to 3.56, we witness a period-doubling bifurcation: Figure 6C shows the system now oscillating over eight points. As we approach the chaotic regime, the range of parameter values in which our system behaves chaotically, the period-doubling bifurcations start to come more quickly. Figure 6D shows that several additional bifurcations occurred between the growth rates of 3.56 and 3.57.

A kind of structure is slowly being revealed across Figure 6, but we can see it much more clearly as we push the growth rate parameter value deep into the chaotic regime. The phase diagram in Figure 7A reveals the system's attractor at a growth rate of 3.9. Figure 7B visualizes 50 different growth rate parameter values between 3.6 and 4, each with its own color. Those rates that exhibit chaos form parabolas, but some gaps exist where the system occasionally settles down into periodic behavior (e.g., in the teal band when the growth rate is set to 3.83; compare this band of periodicity with Figure 4).

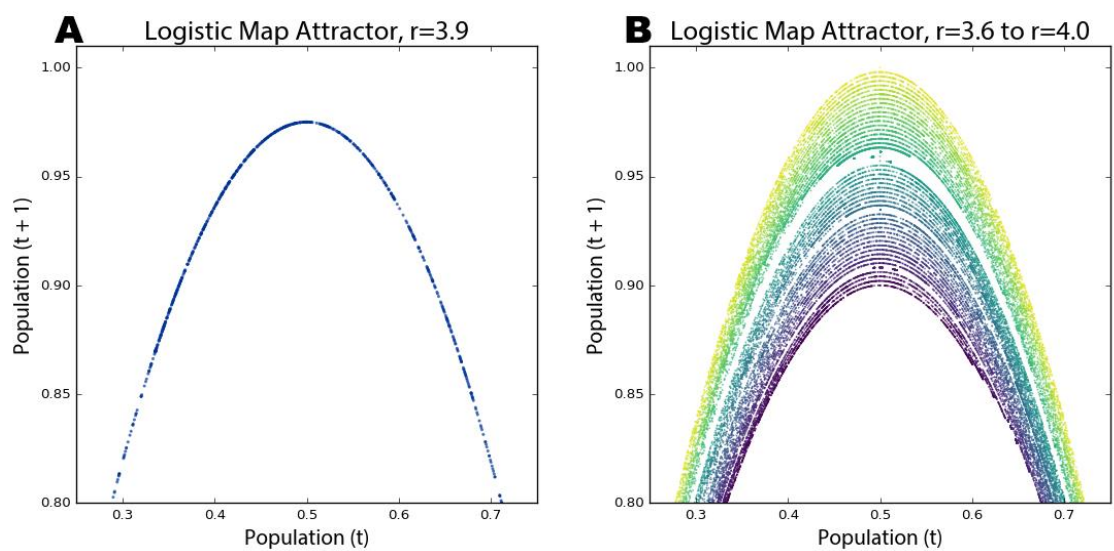

**Figure 7.** Cropped phase diagrams of the logistic map over 200 generations for: (**A**) a growth rate parameter value of 3.9; and (**B**) 50 growth rate parameter values between 3.6 and 4 (the chaotic regime), each with its own colored line

Strange attractors are revealed by these shapes as the system is somehow oddly constrained, yet it never settles into a fixed point or limit cycle like it did in Figure 6. Instead it just bounces around different population values (i.e., points on the parabola) forever without ever repeating the same value twice. It is impossible to predict if any two consecutive observations appear near each other or far apart on the parabola. Further, the parabolas in Figure 7B never overlap due to their fractal geometry and the deterministic nature of the logistic map. Consider: if two different parameter values could ever land on the exact same point, their systems would have to evolve identically over time because the logistic map is deterministic. We can see in these visualizations that this indeed never happens. While the dynamics of a chaotic system appear to have no pattern whatsoever, in reality, they conform to a remarkable fractal pattern, a strange attractor, which confines the system to a limited slice of state space and ensures that no state will ever repeat [62]. Fractals are indeed strange. Rather than having a whole-number dimension such as two or three, they are characterized by a fractional (hence, fractal) dimension [55,61,63]. The fractal dimension refers to the space-filling characteristics of a curve that, through self-similarity, becomes a bit more than a one-dimensional line yet a bit less than a two-dimensional plane.

These visualizations have all plotted quantitative data to better explain and understand the qualitative behavior of a nonlinear dynamical system. A cobweb plot is a visualization technique particularly well-suited to revealing the qualitative behavior of one-dimensional maps, allowing us to analyze the long-term evolution of such systems under recursive iteration [37]. The cobweb plots drawn by Pynamical in Figure 8 consist of three lines: a diagonal gray identity line representing $y = x$, a red curve representing the logistic map as $y = f(x)$ for a given parameter value and a blue line tracing the path of the cobweb. To draw a cobweb:

1.  Begin on the *x*-axis at the point $(x, 0)$ where $x$ is the initial population value (0.5 in our example), and draw a vertical line to the red function curve; this new point is at $(x, f(x))$.
2.  Draw a horizontal line from this point to the gray identity line; this new point is at $(f(x), f(x))$.
3.  Draw a vertical line from this point to the red function curve; this new point is at $(f(x), f(f(x)))$.
4.  Repeat Steps 2 and 3 recursively. The cobwebs in Figure 8 were iterated 100 times.

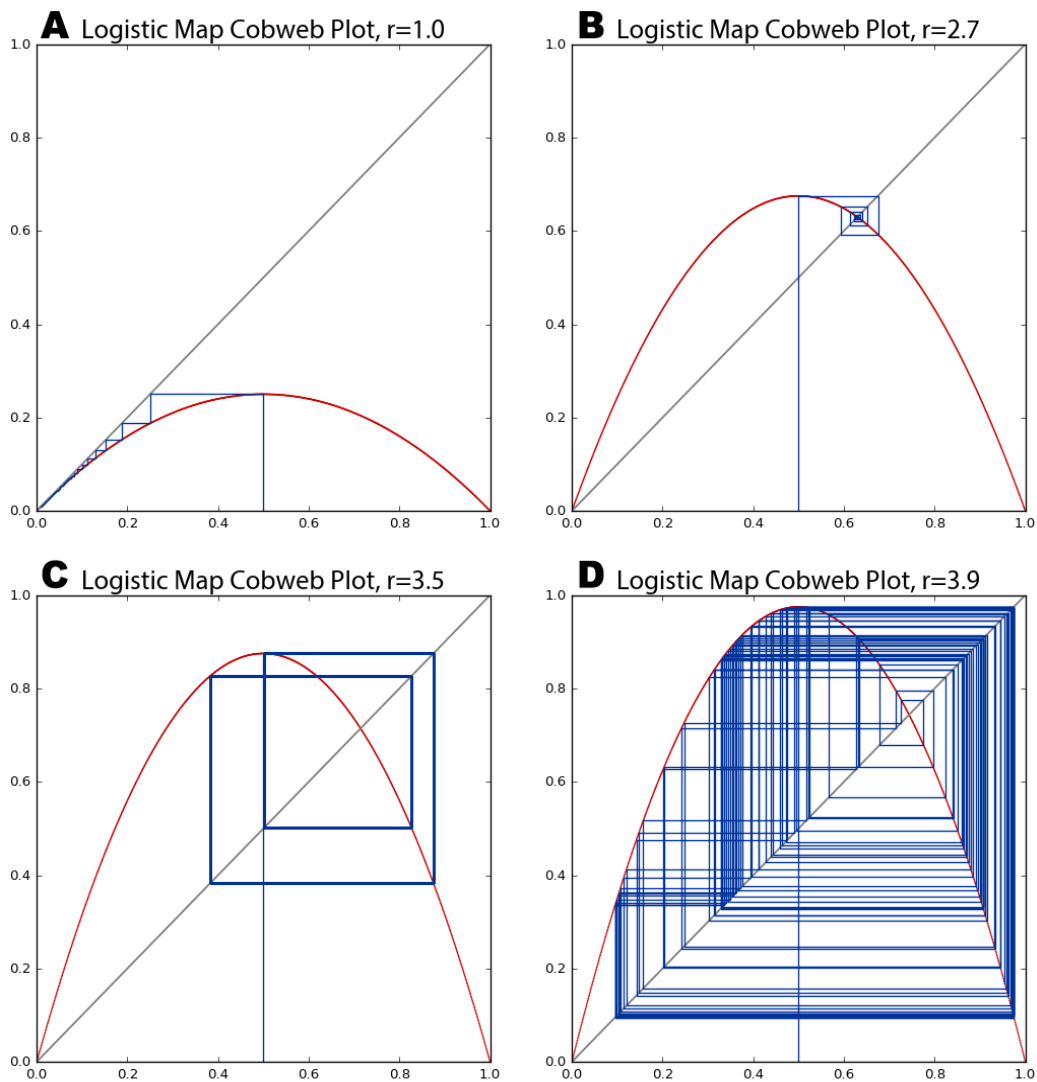

**Figure 8.** Cobweb plots of the logistic map for growth rate parameter values of: (**A**) 1; (**B**) 2.7; (**C**) 3.5; (**D**) 3.9. The diagonal gray identity line represents $y = x$; the red curve represents the logistic map as $y = f(x)$ for each of the four parameter values; and the blue cobweb line represents the system's trajectory over 100 generations.

The blue lines intersect the red curve at those values our system lands on as it iterates from an initial population value of 0.5. In Figure 8A,B, the cobweb shows the system homing in on fixed-point attractors of zero and 0.65, respectively. At a growth rate of 3.5 (Figure 8C), the system oscillates over four points in its limit cycle attractor, denoted by rectangular closed loops. The points where the blue lines intersect the red curve are the same as those revealed by the attractor in Figure 6B for the same parameter value. Finally, Figure 8D visualizes our system's behavior in the chaotic regime at a growth rate of 3.9. The chaotic orbit fills the plot with rectangles, an eventually infinite number of never-repeating trajectories that form a fractal cobweb throughout the diagram.

## 5. Chaos and Randomness

Phase diagrams are useful for visually revealing strange attractors in time series data, like that produced by the logistic map, because they embed this one-dimensional data into a two- or even three-dimensional state space. It can be difficult to ascertain if certain time series are deterministic or just random if we do not fully understand their underlying dynamics [64]. Take the two series plotted by Pynamical in Figure 9 as an example. Both of the lines seem to jump around randomly. The red line does depict random data, but the blue line comes from our logistic model when the growth rate is set to 3.99. This is deterministic chaos, but it is difficult to differentiate from randomness. Instead in Figure 10, we visualize these same two datasets with phase diagrams rather than time graphs, giving us a clear window into the qualitative behavior of our systems. Now, we can clearly see our chaotic system constrained by its strange attractor. By contrast, the random data just look like the noise that they actually are.

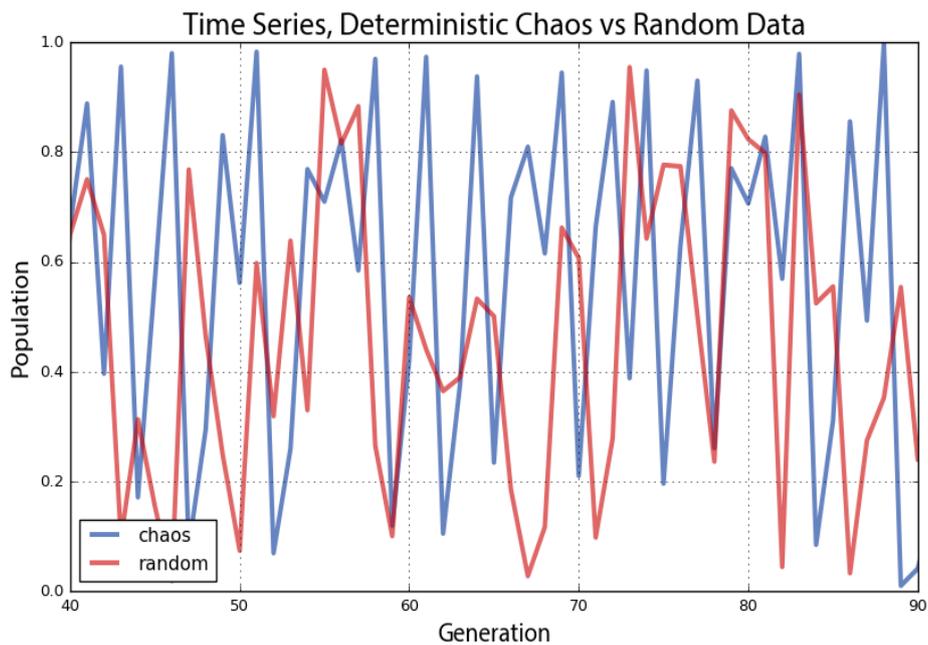

**Figure 9.** Plot of two time series, one deterministic/chaotic from the logistic map (**blue**), and one random (**red**).

This is particularly revealing in a three-dimensional phase diagram from Pynamical (Figure 10B) that embeds our time series into a three-dimensional state space by plotting the population value at generation $t + 2$ versus the value at $t + 1$ versus the value at $t$. This plot essentially extrudes our two-dimensional plot (Figure 10A), then pans and rotates the viewpoint. In fact, if we looked straight down at the *xy*-plane of the three-dimensional plot in Figure 10B, it would look identical to the two-dimensional plot in Figure 10A (see Appendix A for an animated visualization of this).

Strange attractors stretch and fold state space in higher dimensions, allowing their fractal forms to fill space without ever producing the same value twice.

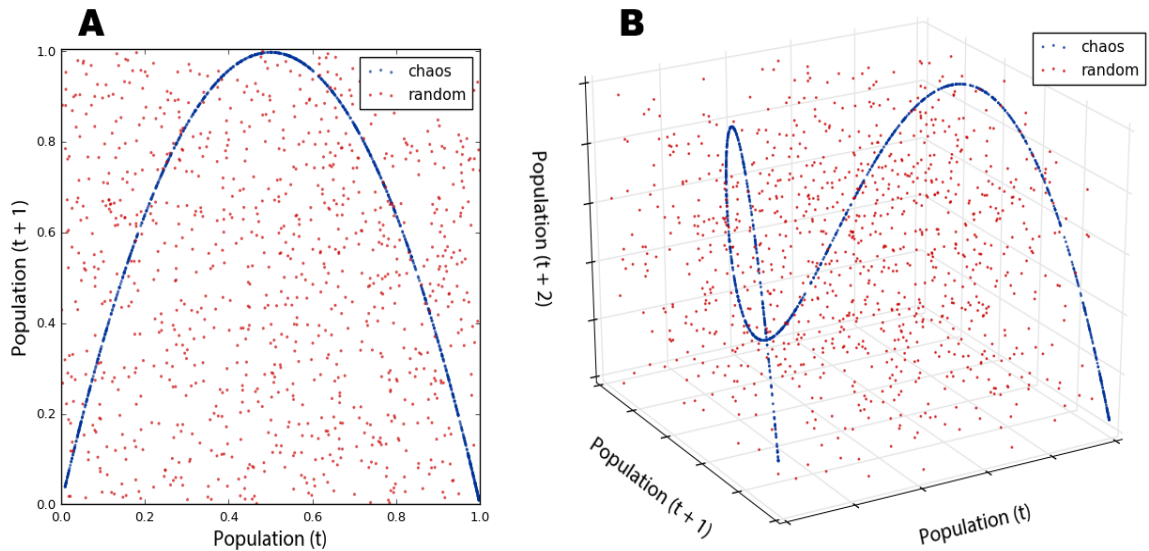

**Figure 10.** Phase diagrams of the time series in Figure 9. (**A**) is a two-dimensional state space version (the *xy*-plane) of the three-dimensional one (**B**).

To press this further, we can use Pynamical to visualize the rest of the logistic map's chaotic regime in three dimensions: the phase diagram in Figure 11 is a three-dimensional version of the two-dimensional state space we saw in Figure 7B. The novel color coding exposes the dynamical system's behavior across the chaotic regime: information virtually impenetrable without visualization. The beautiful structure of the strange attractor is revealed as it twists and curls around its three-dimensional state space (see Appendix A for an animated visualization). This structure again demonstrates that our apparently random time series data from the logistic model is not truly random at all. Instead, it is aperiodic deterministic chaos, constrained by a mind-bending strange attractor. No matter how much we zoom in, the parabolas never overlap and no point ever repeats itself.

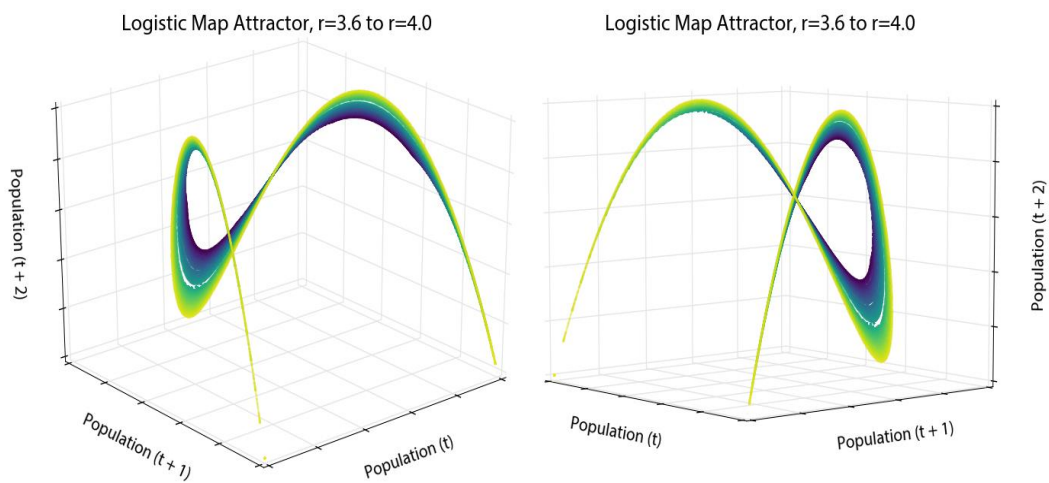

**Figure 11.** Two different viewing perspectives of a single three-dimensional phase diagram of the logistic map over 200 generations for 50 growth rate parameter values between 3.6 and 4, each with its own colored line.

## 6. Unpredictable Systems: The Butterfly Effect

Attractors have a basin of attraction: a set of points that the system's dynamics will pull into this attractor over time [65]. This is easily seen with a cobweb plot. Figure 12 shows how the logistic map's basin of attraction (when the growth rate is 2.7) pulls three different initial population values into the same fixed-point attractor. The initial state of the system will eventually become unknowable, because any one of many different possible points in the basin of attraction could have been the one pulled into the attractor.

By contrast, chaotic systems are characterized by their sensitive dependence on initial conditions. Their strange attractors are globally stable, yet locally unstable: they have basins of attraction, yet within a strange attractor, infinitesimally close points diverge over time without ever leaving the attractor's confines. This divergence can be measured by Lyapunov exponents [66], the calculation of which is described by Wolf et al. [67]. If the Lyapunov exponent's value is positive, then the two points move apart over time at an exponential rate. If the Lyapunov exponent is negative, then these points converge exponentially quickly, such as toward a fixed point or limit cycle. Finally, the Lyapunov exponent is zero when there is a bifurcation [68]. For example, with our logistic model, the Lyapunov exponent is zero when the growth rate is set to one or three because they are bifurcation points; it is negative for most growth rates, such as $0 \leq r < 1$ and $1 < r < 3$, because they have fixed-point or limit cycle attractors; and it is positive for the chaotic regime (exclusive of those occasional windows when the system resumes brief periodicity, such as when the growth rate is 3.83). A positive Lyapunov exponent indicates that the system has a highly sensitive dependence on initial conditions and is a common signature of chaos [69–71].

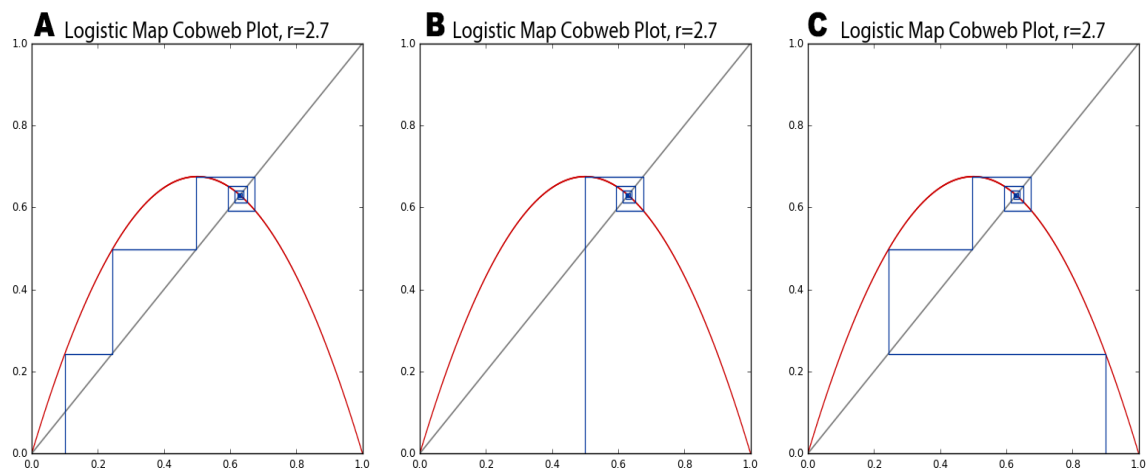

**Figure 12.** Cobweb plots of the logistic map pulling initial population values of 0.1 (**A**), 0.5 (**B**) and 0.9 (**C**) into the same fixed-point attractor over time. At this growth rate parameter value of 2.7, the Lyapunov exponent is negative.

This nonlinear divergence of very similar values makes real-world modeling and prediction difficult, because we must measure the parameters and system state with infinite precision. Otherwise, tiny errors in measurement or rounding are compounded over time until the system eventually diverges drastically from the prediction. In the real world, infinite precision is impossible. It was through one such rounding error that Lorenz first discovered chaos. Recall his words at the beginning of this article: "the present determines the future, but the approximate present does not approximately determine the future" [39].

As a demonstration of this, we run the logistic model with two very similar initial population values, shown in Figure 13. Both have the same growth rate parameter value of 3.9. The blue line represents an initial population value of 0.5, and the red line represents an initial population of 0.50001. These two initial conditions are extremely close to one another, and accordingly, their trajectories look essentially identical for the first 30 generations. After that, however, the minuscule difference in initial conditions compounds to the point that by the 40th generation, the two lines show little in common. What began as nearly indistinguishable initial conditions produces completely different outcomes over time due to nonlinearity and exponential divergence.

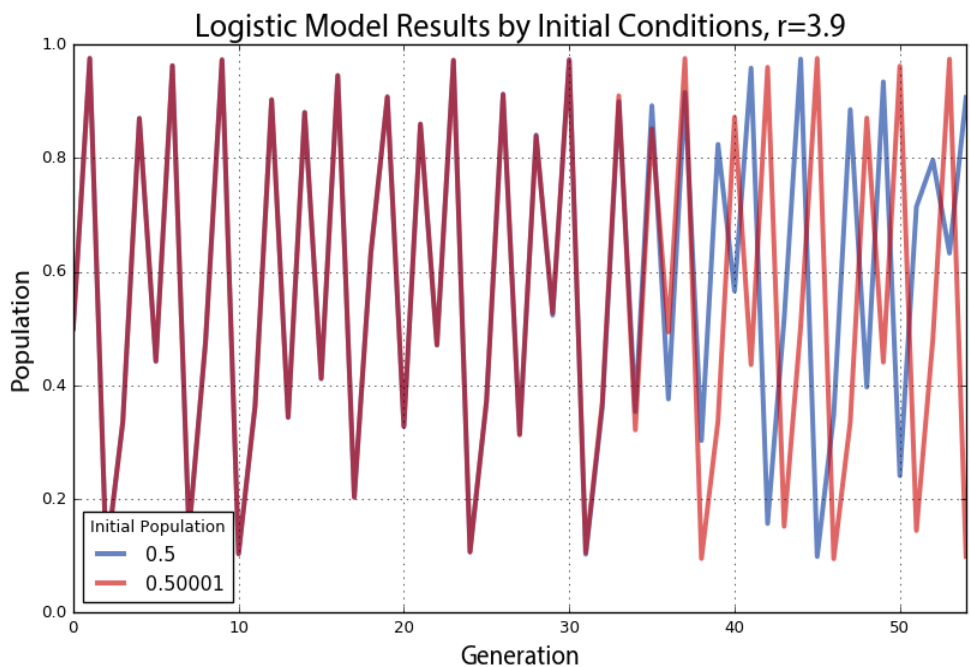

**Figure 13.** Plot of two time series with identical dynamics, one starting at an initial population value of 0.5 (**blue**) and the other starting at 0.50001 (**red**). At this growth rate parameter value of 3.9, the Lyapunov exponent is positive; thus, the system is chaotic, and we can see the nearby points diverge over time.

If our knowledge of these two systems began at Generation 50, we would have no way of guessing that they were nearly identical in the beginning. With chaos, history is thus lost to time, and prediction of the future is only as accurate as our measurements. Human measurements are never infinitely precise, so in real-world chaotic systems, errors compound, and the future becomes entirely unknowable given long enough time horizons. This phenomenon is famously known as the butterfly effect: a butterfly flaps its wings in China and sets off a tornado in Texas. Small events compound and irreversibly alter the future of the universe. In Figure 13, a tiny fluctuation of 0.00001 makes an enormous difference in the behavior and state of the system 40 generations later. Although this system's future cannot be predicted, we can characterize its dynamics geometrically with phase diagrams, bifurcation plots and cobweb plots; and statistically with Lyapunov exponents and fractal dimensions.

## 7. Conclusions

This article had three aims. First, it argued for the usefulness of several visualization methods to analyze and understand the behavior of nonlinear dynamical systems. Second, it introduced the foundational concepts of nonlinear dynamics, chaos, fractals, self-similarity and the limits of prediction. It argued that information visualization is a key way to engage with these system concepts. Third, and by way of the first two, it demonstrated the Pynamical software package for visualizing the behavior of discrete nonlinear dynamical systems. In particular, Pynamical translates these concepts and tools into the Python programming language because it is prevalent, fast and easy to use (details in Appendix A). This package fills a need for a free, fast, simple, extensible tool to introduce and analyze nonlinear dynamical systems' behavior visually, making it a useful platform for research, engineering and pedagogy.

Nonlinear systems are extremely difficult to solve analytically because they cannot be broken down into constituent parts. Instead of solving them, this article presented visual geometric methods to explore system behavior and stability. These methods have broad applicability to scholarly and professional visual data analysis by revealing hidden structure and patterns in time series whose underlying dynamics may not be well known. In particular, they reveal the qualitative behavior of nonlinear dynamical systems over time and/or in response to parameter variations.

This article used the logistic map to define such a set of nonlinear dynamics. As simple as this model was, at different growth rate parameter values, it produced stability, periodic oscillations or chaos. We used Pynamical to create bifurcation diagrams and cobweb plots to visualize this behavior across different parameter values. In the chaotic regime, the system jumped seemingly randomly between all population values. Accordingly, we used Pynamical to embed the data into higher-dimensional state space to create phase diagrams to visualize the system's strange attractor and understand its constrained, deterministic dynamics. Finally, we explored the butterfly effect's implications of nonlinearity on system sensitivity, as infinitesimal differences in initial conditions compounded over time until nearly identical systems had diverged drastically. Thus, in many nonlinear systems, there are fundamental limits to knowledge and prediction. These visualization methods can help researchers discover, examine and understand the behavior of nonlinear dynamical systems and chaos.

**Conflicts of Interest:** The authors declare no conflict of interest.

## Appendix A

Pynamical and all of the code used to develop these models and produce these visualizations are available in a public repository on GitHub at https://github.com/gboeing/pynamical. Pynamical is built on top of Python's numpy, numba, pandas and matplotlib code libraries:

- numpy is a numerical library that handles the underlying data vectors;
- numba provides just-in-time compilation for optimized performance;
- pandas handles the higher-level data structures and analysis;
- matplotlib is the engine used to produce the visualizations and graphics.

Python was selected for developing Pynamical and the visualizations in this article because it is a standard programming environment for information visualization. It offers three key advantages: prevalence, speed and ease of use. First, Python has become a widely-prevalent language. The TIOBE index, a common measure of programming language popularity, ranks Python as the fifth most popular programming language as of 2016. In contrast, other languages commonly used for data science, visualization and mathematics are ranked much lower (e.g., R, MATLAB and SAS are only ranked 18th, 19th and 21st, respectively, in popularity). Developing tools for exploring, understanding and visualizing dynamical systems in Python makes them available to a much wider audience of systems analysts, researchers and students. Furthermore, unlike many languages, Python is completely free and open-source. It is also multi-purpose, and researchers can use its standard syntax and grammar for everything from statistical modeling, to cartography, to full software development. Becoming a Swiss army knife of the computational science world, Python has grown popular and powerful. Today, innumerable researchers and developers contribute open-source libraries of pre-packaged functionality for all to repurpose.

Second, Pynamical is fast. It is built on top of Python's pandas, numpy and numba libraries for quick vectorized numerical computation. Although it is an interpreted language, these libraries provide compiled components that make mathematical modeling fast and efficient and can take advantage of the optimized mathematical routines of Intel processors' Math Kernel Library. Pynamical also takes advantage of just-in-time code compilation to offer additional speed-ups. Third, Python is very easy to use, and this simplicity translates to Pynamical. Most of the visualizations in this article can be created from scratch in just one or two lines of straightforward code. The following examples demonstrate this simplicity. The user merely imports Pynamical into the Python environment, then runs the following code to produce the visualization:

- Figure 2: bifurcation_plot(simulate(num_rates = 1000))
- Figure 4: bifurcation_plot(simulate(rate_min = 3.7, rate_max = 3.9, num_rates = 1000))
- Figure 6D: phase_diagram(simulate(num_gens = 100, rate_min = 3.57))
- Figure 8D: cobweb_plot(r = 3.9, x0 = 0.5)
- Figure 11: phase_diagram_3d(simulate(num_gens = 4000, rate_min = 3.6, num_rates = 50)).

Pynamical defines extensible functions to express a discrete map's equation and encapsulate a model to simulate it iteratively. The logistic map, the Singer map and the cubic map are built-in by default, but any other iterated map can be defined and added by the user. Pynamical also defines a function to convert model output into *x-y* points, as well as functions to plot these points as a bifurcation diagram, a cobweb plot, an animated cobweb plot, a two-dimensional phase diagram, a three-dimensional phase diagram and an animated three-dimensional phase diagram. Animated cobweb plots of the entire parameter space and animated three-dimensional phase diagrams extending those presented in this study are also available in the GitHub repository. They shed particular light on the fractal nature of strange attractors as they stretch and fold state space, thus serving as an indispensable tool for pedagogy and visual information presentation.

## References

1. Hastings, A.; Hom, C.L.; Ellner, S.; Turchin, P.; Godfray, H.C.J. Chaos in Ecology: Is Mother Nature a Strange Attractor? *Annu. Rev. Ecol. Syst.* **1993**, *24*, 1–33. [CrossRef]
2. Rickles, D.; Hawe, P.; Shiell, A. A Simple Guide to Chaos and Complexity. *J. Epidemiol. Commun. Health* **2007**, *61*, 933–937. [CrossRef] [PubMed]
3. Suetani, H.; Soejima, K.; Matsuoka, R.; Parlitz, U.; Hata, H. Manifold Learning Approach for Chaos in the Dripping Faucet. *Phys. Rev. E* **2012**, *86*, 036209. [CrossRef] [PubMed]
4. Singh, S.L.; Mishra, S.N.; Sinkala, W. A New Iterative Approach to Fractal Models. *Commun. Nonlinear Sci. Numer. Simul.* **2012**, *17*, 521–529. [CrossRef]
5. Hoshi, R.A.; Pastre, C.M.; Vanderlei, L.C.M.; Godoy, M.F. Poincaré Plot Indexes of Heart Rate Variability: Relationships with Other Nonlinear Variables. *Auton. Neurosci.* **2013**, *177*, 271–274. [CrossRef] [PubMed]
6. Babbs, C.F. Initiation of Ventricular Fibrillation by a Single Ectopic Beat in Three Dimensional Numerical Models of Ischemic Heart Disease: Abrupt Transition to Chaos. *J. Clin. Exp. Cardiol.* **2014**, *5*, 2–11. [CrossRef]
7. Glass, L. Introduction to Controversial Topics in Nonlinear Science: Is the Normal Heart Rate Chaotic? *Chaos* **2009**, *19*, 028501. [CrossRef] [PubMed]
8. Hong, Z.; Dong, J. Chaos Theory and Its Application in Modern Cryptography. In Proceedings of the 2010 International Conference on Computer Application and System Modeling (ICCASM 2010), Taiyuan, China, 22–24 October 2010; pp. 332–334.
9. Makris, G.; Antoniou, I. Cryptography with Chaos. In Proceedings of the 5th Chaotic Modeling and Simulation International Conference, Athens, Greece, 12–15 June 2012; pp. 309–318.
10. Guastello, S.J. *Chaos, Catastrophe, and Human Affairs: Applications of Nonlinear Dynamics to Work, Organizations, and Social Evolution*; Psychology Press: New York, NY, USA, 2013.
11. Richards, D. From Individuals to Groups: The Aggregation of Votes and Chaotic Dynamics. In *Chaos Theory in the Social Sciences*; Kiel, L.D., Elliott, E., Eds.; University of Michigan Press: Ann Arbor, MI, USA, 1996; pp. 89–116.

12. Batty, M.; Longley, P. *Fractal Cities: A Geometry of Form and Function*; Academic Press: London, UK, 1994.
13. Batty, M.; Xie, Y. Self-Organized Criticality and Urban Development. *Discret. Dyn. Nat. Soc.* **1999**, *3*, 109–124. [CrossRef]
14. Benguigui, L.; Czamanski, D.; Marinov, M.; Portugali, Y. When and Where is a City Fractal? *Environ. Plan. B* **2000**, *27*, 507–519. [CrossRef]
15. Shen, G. Fractal Dimension and Fractal Growth of Urbanized Areas. *Int. J. Geogr. Inf. Sci.* **2002**, *16*, 419–437. [CrossRef]
16. Chen, Y.; Zhou, Y. Scaling Laws and Indications of Self-Organized Criticality in Urban Systems. *Chaos Solitons Fractals* **2008**, *35*, 85–98. [CrossRef]
17. Chen, W.C. Nonlinear Dynamics and Chaos in a Fractional-Order Financial System. *Chaos Solitons Fractals* **2008**, *36*, 1305–1314. [CrossRef]
18. Guégan, D. Chaos in Economics and Finance. *Annu. Rev. Control* **2009**, *33*, 89–93. [CrossRef]
19. Puu, T. *Attractors, Bifurcations, & Chaos: Nonlinear Phenomena in Economics*, 2nd ed.; Springer Science & Business Media: New York, NY, USA, 2013.
20. Rosser, J.B. Chaos Theory and Rationality in Economics. In *Chaos Theory in the Social Sciences*; Kiel, L.D., Elliott, E., Eds.; University of Michigan Press: Ann Arbor, MI, USA, 1996; pp. 199–213.
21. Oxley, L.; George, D.A.R. Economics on the Edge of Chaos: Some Pitfalls of Linearizing Complex Systems. *Environ. Model. Softw.* **2007**, *22*, 580–589. [CrossRef]
22. Hamouche, M.B. Can Chaos Theory Explain Complexity In Urban Fabric? Applications in Traditional Muslim Settlements. *Nexus Netw. J.* **2009**, *11*, 217–242. [CrossRef]
23. Ostwald, M.J. The Fractal Analysis of Architecture: Calibrating the Box-Counting Method Using Scaling Coefficient and Grid Disposition Variables. *Environ. Plan B* **2013**, *40*, 644–663. [CrossRef]
24. Cartwright, T.J. Planning and Chaos Theory. *J. Am. Plan. Assoc.* **1991**, *57*, 44–56. [CrossRef]
25. Innes, J.E.; Booher, D.E. *Planning with Complexity*; Routledge: London, UK, 2010.
26. Batty, M.; Marshall, S. The Origins of Complexity Theory in Cities and Planning. In *Complexity Theories of Cities Have Come of Age*; Portugali, J., Meyer, H., Stolk, E., Tan, E., Eds.; Springer: Berlin, Germany, 2012; pp. 21–45.
27. Batty, M. *The New Science of Cities*; MIT Press: Cambridge, MA, USA, 2013.
28. Alpigini, J.J. Dynamical System Visualization and Analysis via Performance Maps. *Inf. Vis.* **2004**, *3*, 271–287. [CrossRef]
29. Layek, G.C. *An Introduction to Dynamical Systems and Chaos*; Springer: New Delhi, India, 2015.
30. Chen, C. *Information Visualization: Beyond the Horizon*, 2nd ed.; Springer: London, UK, 2006.
31. Lorenz, E.N. Deterministic Nonperiodic Flow. *J. Atmos. Sci.* **1963**, *20*, 130–141. [CrossRef]
32. May, R.M. Simple Mathematical Models with Very Complicated Dynamics. *Nature* **1976**, *261*, 459–467. [CrossRef] [PubMed]
33. Packard, N.H.; Crutchfield, J.P.; Farmer, J.D.; Shaw, R.S. Geometry from a Time Series. *Phys. Rev. Lett.* **1980**, *45*, 712–716. [CrossRef]
34. Bradley, E. Time Series Analysis. In *Intelligent Data Analysis: An Introduction*, 2nd ed.; Hand, D., Berthold, M., Eds.; Springer: Berlin, Germany, 2003.
35. Bradley, E.; Kantz, H. Nonlinear Time-Series Analysis Revisited. *Chaos* **2015**, *25*, 097610. [CrossRef] [PubMed]
36. Chettiparamb, A. Metaphors in Complexity Theory and Planning. *Plan Theory* **2006**, *5*, 71–91. [CrossRef]
37. Tomida, A.G. Matlab Toolbox and GUI for Analyzing One-Dimensional Chaotic Maps. In Proceedings of the 2008 International Conference on Computational Sciences and Its Applications, Perugia, Italy, 30 June–3 July 2008; pp. 321–330.
38. Stewart, I. The Lorenz Attractor Exists. *Nature* **2000**, *406*, 948–949. [CrossRef] [PubMed]
39. Danforth, C.M. Chaos in an Atmosphere Hanging on a Wall. Available online: http://mpe2013.org/2013/03/17/chaos-in-an-atmosphere-hanging-on-a-wall/ (accessed on 1 September 2016).
40. Gleick, J. *Chaos: Making a New Science*; Cardinal: Indianapolis, IN, USA, 1991.
41. May, R.M. Biological Populations with Nonoverlapping Generations: Stable Points, Stable Cycles, and Chaos. *Science* **1974**, *186*, 645–647. [CrossRef] [PubMed]
42. Li, W.; Wang, K.; Su, H. Optimal Harvesting Policy for Stochastic Logistic Population Model. *Appl. Math. Comput.* **2011**, *218*, 157–162. [CrossRef]

43. Pastijn, H. Chaotic Growth with the Logistic Model of P.F. Verhulst. In *The Logistic Map and the Route to Chaos*; Ausloos, M., Dirickx, M., Eds.; Springer: Berlin, Germany, 2006; pp. 3–11.

44. Strogatz, S.H. *Nonlinear Dynamics and Chaos*, 2nd ed.; Westview Press: Boulder, CO, USA, 2014.

45. Ruelle, D.; Takens, F. On the Nature of Turbulence. *Commun. Math. Phys.* **1971**, *20*, 167–192. [CrossRef]

46. Shilnikov, L. Bifurcations and Strange Attractors. *Proc. Int. Congr. Math.* **2002**, *3*, 349–372.

47. Grebogi, C.; Ott, E.; Yorke, J.A. Chaos, Strange Attractors, and Fractal Basin Boundaries in Nonlinear Dynamics. *Science* **1987**, *238*, 632–638. [CrossRef] [PubMed]

48. Gershenson, C. Introduction to Chaos in Deterministic Systems. Available online: http://arxiv.org/abs/nlin/0308023 (accessed on 10 November 2016).

49. Wu, G.C.; Baleanu, D. Discrete Fractional Logistic Map and Its Chaos. *Nonlinear Dyn.* **2014**, *75*, 283–287. [CrossRef]

50. Li, T.Y.; Yorke, J.A. Period Three Implies Chaos. *Am. Math. Mon.* **1975**, *82*, 985–992. [CrossRef]

51. Feigenbaum, M.J. Quantitative Universality for a Class of Nonlinear Transformations. *J. Stat. Phys.* **1978**, *19*, 25–52. [CrossRef]

52. Feigenbaum, M.J. Universal Behavior in Nonlinear Systems. *Phys. Nonlinear Phenom.* **1983**, *7*, 16–39. [CrossRef]

53. Hénon, M. A Two-Dimensional Mapping with a Strange Attractor. *Commun. Math. Phys.* **1976**, *50*, 69–77. [CrossRef]

54. Farmer, J.D.; Ott, E.; Yorke, J.A. The Dimension of Chaotic Attractors. *Phys. Nonlinear Phenom.* **1983**, *7*, 153–180. [CrossRef]

55. Grassberger, P.; Procaccia, I. Characterization of Strange Attractors. *Phys. Rev. Lett.* **1983**, *50*, 346–349. [CrossRef]

56. Mandelbrot, B.B. How Long Is the Coast of Britain? *Science* **1967**, *156*, 636–638. [CrossRef] [PubMed]

57. Mandelbrot, B.B. *The Fractal Geometry of Nature*; Macmillan: New York, NY, USA, 1983.

58. Mandelbrot, B.B. *Multifractals and 1/f Noise*; Springer: New York, NY, USA, 1999.

59. Huikuri, H.V.; Mäkikallio, T.H.; Peng, C.K.; Goldberger, A.L.; Hintze, U.; Møller, M. Diamond Study Group. Fractal Correlation Properties of RR Interval Dynamics and Mortality in Patients with Depressed Left Ventricular Function after an Acute Myocardial Infarction. *Circulation* **2000**, *101*, 47–53. [CrossRef] [PubMed]

60. Takens, F. Detecting Strange Attractors in Turbulence. In *Dynamical Systems and Turbulence*; Rand, D., Young, L.S., Eds.; Springer: Berlin, Germany, 1981; pp. 366–381.

61. Theiler, J. Estimating Fractal Dimension. *J. Opt. Soc. Am. A* **1990**, *7*, 1055–1073. [CrossRef]

62. Kekre, H.B.; Sarode, T.; Halarnkar, P.N. A Study of Period Doubling in Logistic Map for Shift Parameter. *Int. J. Eng. Trends Technol.* **2014**, *13*, 281–286. [CrossRef]

63. Clarke, K.C. Computation of the Fractal Dimension of Topographic Surfaces Using the Triangular Prism Surface Area Method. *Comput. Geosci.* **1986**, *12*, 713–722. [CrossRef]

64. Sander, E.; Yorke, J.A. The Many Facets of Chaos. *Int. J. Bifurc. Chaos* **2015**, *25*, 1530011. [CrossRef]

65. Sprott, J.C.; Xiong, A. Classifying and Quantifying Basins of Attraction. *Chaos* **2015**, *25*, 083101. [CrossRef] [PubMed]

66. Brown, T.A. Measuring Chaos Using the Lyapunov Exponent. In *Chaos Theory in the Social Sciences*; Kiel, L.D., Elliott, E., Eds.; University of Michigan Press: Ann Arbor, MI, USA, 1996; pp. 53–66.

67. Wolf, A.; Swift, J.B.; Swinney, H.L.; Vastano, J.A. Determining Lyapunov Exponents from a Time Series. *Phys. Nonlinear Phenom.* **1985**, *16*, 285–317. [CrossRef]

68. Dingwell, J.B. Lyapunov Exponents. In *Wiley Encyclopedia of Biomedical Engineering*; Akay, M., Ed.; John Wiley & Sons: Hoboken, NJ, USA, 2006.

69. Chan, K.S.; Tong, H. *Chaos: A Statistical Perspective*; Springer Science & Business Media: New York, NY, USA, 2013.

70. Hunt, B.R.; Ott, E. Defining Chaos. *Chaos* **2015**, *25*, 097618. [CrossRef] [PubMed]

71. Kantz, H.; Radons, G.; Yang, H. The Problem of Spurious Lyapunov Exponents in Time Series Analysis and Its Solution by Covariant Lyapunov Vectors. *J. Phys. Math. Theor.* **2013**, *46*, 254009. [CrossRef]